\documentclass{article}
\usepackage{ijcai23}
\usepackage{subfigure}
\usepackage{times}
\usepackage{soul}
\usepackage{multirow}
\usepackage{color}
\usepackage{diagbox}
\usepackage{url}
\usepackage[hidelinks]{hyperref}
\usepackage[utf8]{inputenc}
\usepackage[small]{caption}
\usepackage{graphicx}
\usepackage{amsmath}
\usepackage{amsthm}
\usepackage{booktabs}
\usepackage{algorithm}
\usepackage{algorithmic}
\usepackage{amsfonts}
\usepackage[switch]{lineno}
\title{All Information is Necessary: Integrating Speech Positive and Negative Information by Contrastive Learning for Speech Enhancement}

\author{
    Xinmeng Xu$^{1}$, Weiping Tu$^{1,2, 3}$\thanks{Corresponding Author}, Chang Han$^{1}$, Yuhong Yang$^{1,3}$
   \affiliations
    $^1$National Engineering Research Center for Multimedia Software, School of Computer Science, \\Wuhan University,  China\\
    $^2$ Hubei Luojia Laboratory, China\\
    $^3$Hubei Key Laboratory of Multimedia and Network Communication Engineering,\\ Wuhan University, China\\
    \emails
    \{xuxinmeng, tuweiping, yangyuhong\}@whu.edu.cn
}
\begin{document}

\maketitle

\begin{abstract}
Monaural speech enhancement (SE) is an ill-posed problem due to the irreversible degradation process. Recent methods to achieve SE tasks rely solely on positive information, e.g., ground-truth speech and speech-relevant features. Different from the above, we observe that the negative information, such as original speech mixture and speech-irrelevant features, are valuable to guide the SE model training procedure. In this study, we propose a SE model that integrates both speech positive and negative information for improving SE performance by adopting contrastive learning, in which two innovations have consisted. (1) We design a collaboration module (CM), which contains two parts, contrastive attention for separating relevant and irrelevant features via contrastive learning and interactive attention for establishing the correlation between both speech features in a learnable and self-adaptive manner. (2) We propose a contrastive regularization (CR) built upon contrastive learning to ensure that the estimated speech is pulled closer to the clean speech and pushed far away from the noisy speech in the representation space by integrating self-supervised models. We term the proposed SE network with CM and CR as CMCR-Net. Experimental results demonstrate that our CMCR-Net achieves comparable and superior performance to recent approaches.
\end{abstract}

\section{Introduction}

Speech enhancement (SE) is frequently used as the preprocessor of several acoustic tasks, e.g., speech recognition \cite{ref1}, speaker verification \cite{ref2}, speaker diarization \cite{ref3}, etc., for separating speech from background interference signals \cite{ref4}. Recently, deep learning networks have shown their promising performance on SE even in highly non-stationary noise environments \cite{ref5,ref6,ref7,ref8}. However, noisy distortion and noise residue unavoidably remained in speech signals processed by these methods, which indicates some indistinguishable patterns between speech and noise signals.

To tackle this issue, some approaches capture the correlation of speech positive and negative information to make the patterns between speech and noise signals more discriminative. However, the definitions of speech positive and negative information are diverse, which can be concretely grouped into two categories as follows:

\textbf{Category 1: Clean speech and noise.} The methods of this category aim to predict speech by using the characteristics of background noises that are used as negative information, in which a separate algorithm is set to estimate the noise signal \cite{ref9,ref10,ref11,ref12}. These methods attempt to incorporate noise information generally by adding constraints to the loss function or by directly predicting noise as prior knowledge. However, they are unsuitable for low signal-to-noise ratio (SNR) and non-stationary noise conditions since the noises are always unstructured and hard to model, which is unable to sufficiently estimate complete noise components \cite{ref13}.

\textbf{Category 2: Speech features and noise features.} The methods of this category extract features of speech as positive information while features of noises as negative information by utilizing a two-branch model to separately model the speech and noise signals. Afterward, several interaction modules are inserted between two branches to leverage information learned from the other branch to enhance the target signal modeling \cite{ref8,ref14}. Consequently, the positive and negative information have interacted in a latent representation space, and the interaction modules make the simultaneous modeling of two signals feasible and effective. Despite that the two-branch approach ensures the integrity of speech and noise information during the extraction process, some overlapped information is inevitably extracted and the two-branch structure needs numerous parameters that bring high cost and high difficulty of network training.

The analysis above motivates us to design a new category to explore the correlation between speech positive and negative information, which can overcome the limitations that are: (1) extraction of redundant and incomplete information by modeling speech and noise, and (2) high cost and high difficulty training process caused by independent speech and noise information modeling. To this end, rather than explore the correlation between speech and noise, we are aiming to explore the correlation between speech relevant and irrelevant features extracted from ``high attention'' and ``low attention'' areas partitioned by attention mechanism. Therefore, only a single-branch structure is required to build the network. In addition, since the attention mechanism is unable to accurately recognize speech and noise components, we adopt clean target speech and noisy input speech to guide the network training to avoid redundant and incomplete information in speech relevant and irrelevant features.

In this study, we propose a new category to make the exploration of the correlation between speech positive and negative information more suitable for SE tasks through contrastive learning by using two core designs. First, we treat the speech-relevant features as positive information and speech-irrelevant features as negative information and propose a collaboration module (CM) that consists of contrastive attention and interactive attention modules. Particularly, the contrastive attention module is designed to separate speech relevant and irrelevant features in the deep representation space by combining self-attention and contrastive learning, and the interactive attention module combines these separated features and explores their correlation in a learnable and self-adaptive manner. Second, we adopt noisy speech to provide negative guidance for the network training and propose a contrastive regularization (CR), which is inspired by contrastive learning and consists of two forces, one pulls the prediction closer to the clean speech, and the other pulls the prediction further away from the noisy speech in the representation space. CR improves the performance for SE without introducing additional computation/parameters during the testing phase since they can be removed for inference.

We term the proposed SE framework as CMCR-Net by leveraging CM and CR into encoder-decoder architecture for superior performance. The main contributions of this paper are summarized as follows:
\begin{itemize}
    \item We introduce CM to involve both speech-relevant and speech-irrelevant features to improve the SE model. CM is able to extract rich representations of features from speech-relevant and speech-irrelevant features.
    
    \item We propose a CR for SE to pull the predicted speech to close to positive samples while pushing predict speech far away from negative samples. This is the first attempt to use contrastive learning for SE tasks, we evaluate the CR on different SE models and get very positive feedback.
   
   \item We propose a novel CMCR-Net to effectively enhance noisy speech by adopting CM and CR. CMCR-Net achieves comparable and superior performance on both STOI and PESQ scores.
\end{itemize} 

\section{Related Works}
\subsection{Monaural Speech Enhancement}
Monaural SE aims at separating clean speech from a single-channel noise-corrupted speech signal. Deep learning approaches to SE represent a significant leap in performance. The first deep learning-based SE models use fully connected neural networks (FCNNs) \cite{ref4}, which achieves more satisfactory performance than conventional approaches, thus, further proving the effectiveness of deep learning-based methods. However, the fully connected layers of FCNNs have difficulty in modeling the temporal structure of a speech signal.

Recently, convolutional neural networks (CNNs) have been utilized for SE tasks and achieve remarkable results, which capture implicit information in the speech signal and maintain a small shift in the frequency domain of speech features within a certain range, thereby coping with the changing of speaker identities and acoustic environments \cite{ref15,ref16}. Furthermore, a convolutional encoder-decoder framework is proposed by \cite{ref17}, which improves the SE performance while maintaining an attractive model size. However, they rely only on positive information, i.e., clean speech samples and speech-relevant features, without considering the noisy samples and the speech irrelevant features.

\begin{figure*}
  \centering
  \includegraphics[width=0.90\linewidth]{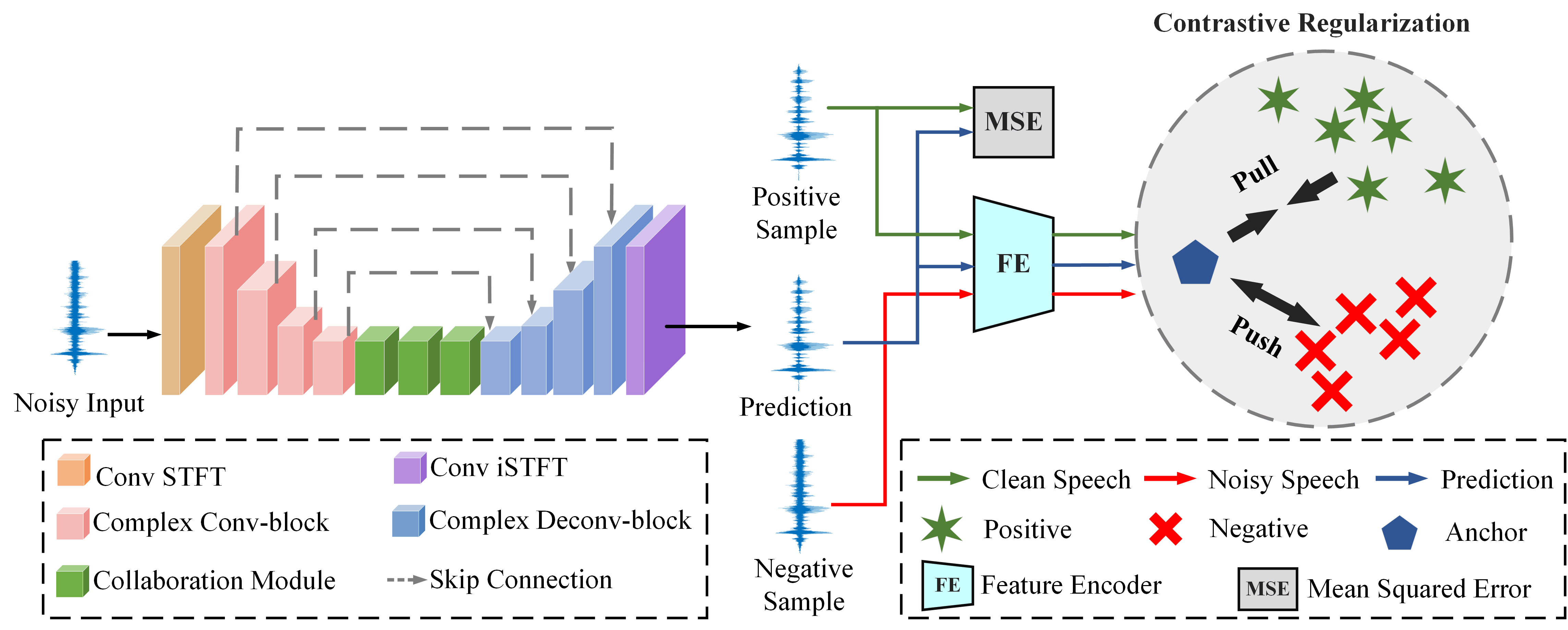}
  \caption{The architecture of the proposed CMCR-Net. It adopts encoder-decoder architecture as the main body. In addition, the loss function of the proposed CMCR-Net consists of MSE loss and contrastive regularization.}
  \label{fig:1}
\end{figure*}

\subsection{Speech Enhancement with Noise Prior}
Rather than only estimating target speech, some SE methods pursue more promising performance by building a noise model for noise prior. Some conventional methods, such as OMLSA \cite{ref10} and spectral subtraction algorithms \cite{ref11}, take this way to achieve noise suppression. However, conventional methods assume the noisy signal is stationary since estimating noise power spectral density of non-stationary noises appears unworkable. To address this issue, Liu et al. \cite{ref12} proposed to achieve spectral subtraction by using DNNs, in which a two-stage paradigm for noise signal estimation and speech signal recovery is introduced. Although DNN is more powerful for modeling noise signals than conventional methods, DNN is only effective for structured noise \cite{ref13}, thus, its generalization capability is limited.

For more accurate noise modeling, Zheng et al. \cite{ref8} proposed a two-branch CNN to predict speech and noise simultaneously and introduce interaction modules at several intermediate layers based on the correlations between predicted speech and the residual signal by subtracting enhanced speech from a noisy signal. However, the two-branch structure results in additional computation/parameters.

\subsection{Contrastive Learning}
Contrastive learning is widely used in self-supervised learning tasks \cite{ref18,ref19,ref20,ref21}. For a given anchor point, contrastive learning aims to pull the anchor close to positive points and push the anchor far away from negative points in the representation space. Contrastive learning often applies to high-level vision tasks, since these tasks are inherently suited for modeling the contrast between positive and negative samples. Besides, Taesung et al. \cite{ref22} have proved that contrastive learning effectively improves unpaired image-to-image translation quality. However, the research on adopting contrastive learning into SE tasks is still a blank area. Conversely, our approach explores applying contrastive learning to SE tasks for superior performance and proposes a novel contrastive loss and a new sampling method for SE.

\section{Proposed Method}
In this section, we present the proposed CMCR-Net model, consisting of two key components, i.e., collaborative module (CM) and contrastive regularization (CR). The detailed architecture of CMCR-Net is shown in Figure~\ref{fig:1}. The CMCR-Net aims at estimating the target speech $\textbf{s}$ from noise-corrupted speech $\textbf{y}$. Besides, CMCR-Net takes noisy speech spectrogram $\textbf{Y}_{r, i}\in \mathbb{R}^{T\times F\times 2}$ as input, where $T$ represents the number of time steps and $F$ represents the number of frequency bands.

Furthermore, CMCR-Net uses convolutional encoder-decoder architecture. Specifically, the encoder consists of 4 complex convolutional layers and the decoder 4 complex deconvolutional layers \cite{ref23}. ELUs are utilized in all convolutional and deconvolutional layers except the output layer since they perform faster convergence and better generalization than ReLUs \cite{ref24}. Batch normalization is inserted between each convolution or deconvolution and activation. Between the encoder and the decoder, 3 CMs are inserted. Additionally, the skip connections are proposed and utilized to concatenate the output of each encoder layer to the input of the corresponding decoder layer.

\begin{figure}[t]
  \centering
  \includegraphics[width=\linewidth]{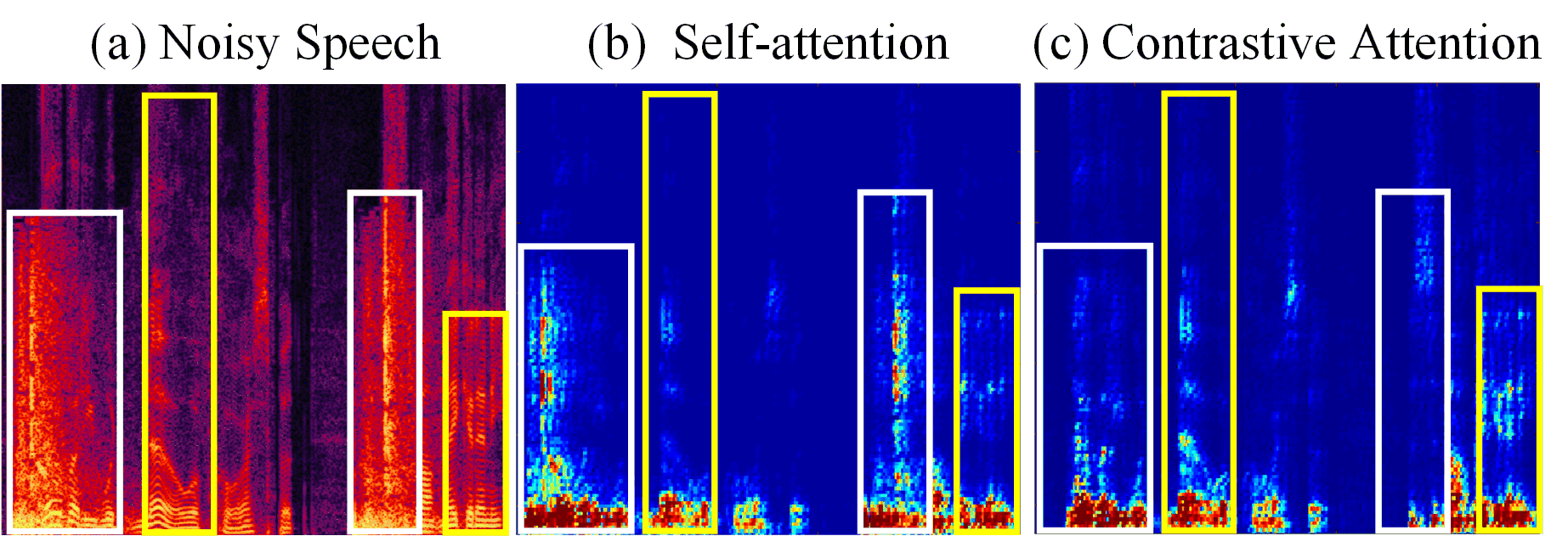}
  \caption{The visualization of attention maps obtained by softmax and inner product in self-attention and contrastive attention, respectively. We can see that self-attention pays attention to speech-irrelevant features and ignore several speech-relevant features, while our contrastive attention can simultaneously leverage speech-relevant features and suppress speech-irrelevant features for SE.}
  \label{fig:2}
\end{figure}

\subsection{Collaboration Module (CM)}
An important concept presented in our work is that speech-relevant information and speech-irrelevant information are all-valuable for SE. Using these two types of information without distinction will bring a passive effect. To address this issue, we propose a collaborative module (CM) by combining speech-relevant and speech-irrelevant features to capture more discriminative information. Note that this combination is not a simple series or parallel connection but is learnable and self-adaptive. The proposed CM consists of two parts: a contrastive attention module and an interactive attention module.

\begin{figure}[t]
  \centering
  \includegraphics[width=\linewidth]{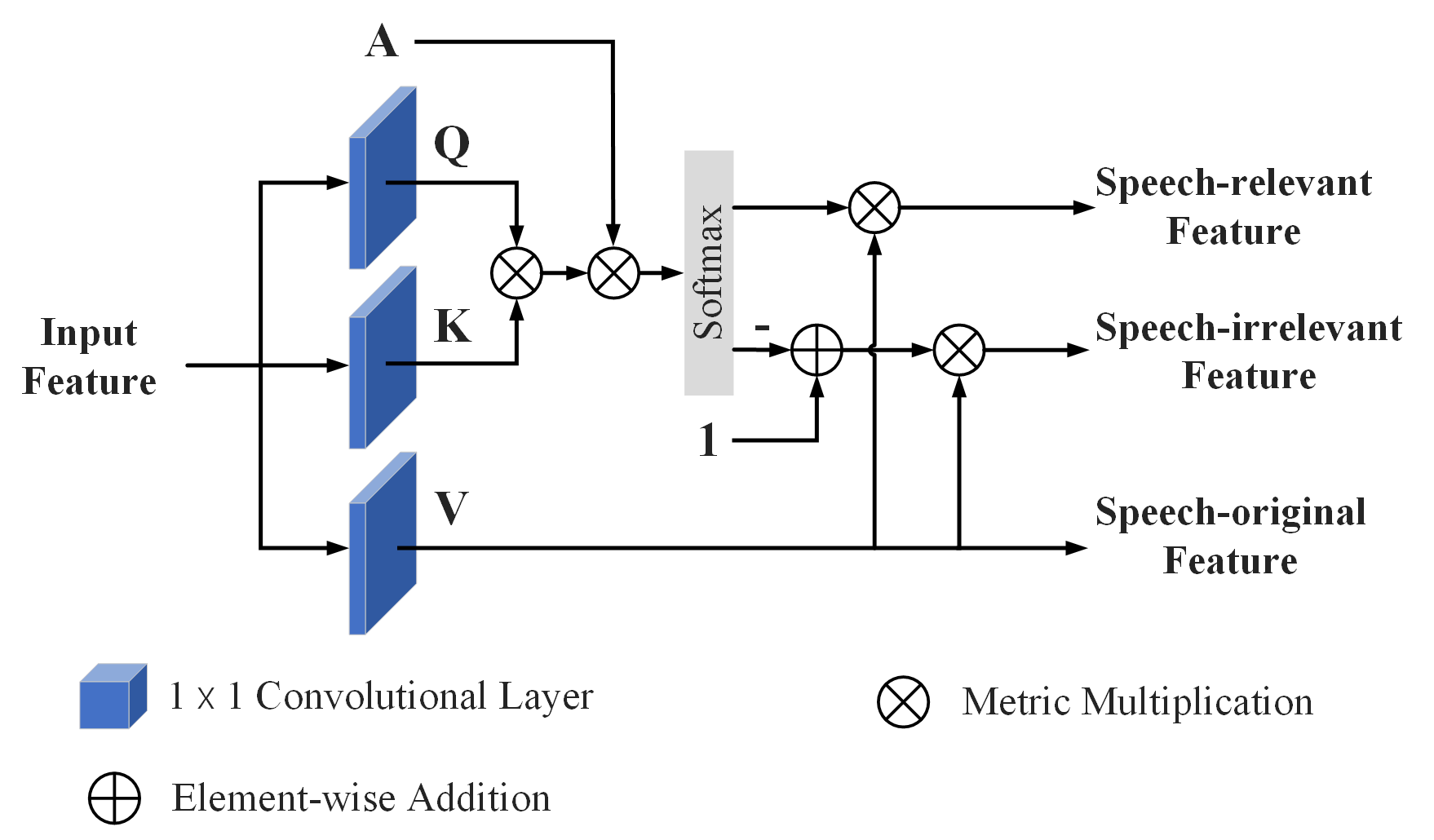}
  \caption{The proposed contrastive attention module.}
  \label{fig:3}
\end{figure}

The contrastive attention module aims to separate the speech relevant and irrelevant features by utilizing self-attention and contrastive learning. Formally, the self-attention module computes the feature mutual similarity among all the spectrogram feature time-frequency locations via metric multiplication operation, in which the regions which have high power are emphasized. In addition, the power of the voice active region is likely to be higher than those of noise ones even in low SNR cases, thus, the self-attention module actually emphasizes voice-active regions where the abundant speech-relevant features are contained. However, in some cases, the noise feature regions have higher power than speech feature regions, and self-attention always highlights the noise feature region while ignoring the speech feature regions, as shown in Figure~\ref{fig:2} (b) where the noise components are highlighted in white boxes and speech components are ignored by self-attention. As a consequence, residual noises and speech distortion may be contained in the estimated speech.

\begin{figure}[t]
  \centering
  \includegraphics[width=\linewidth]{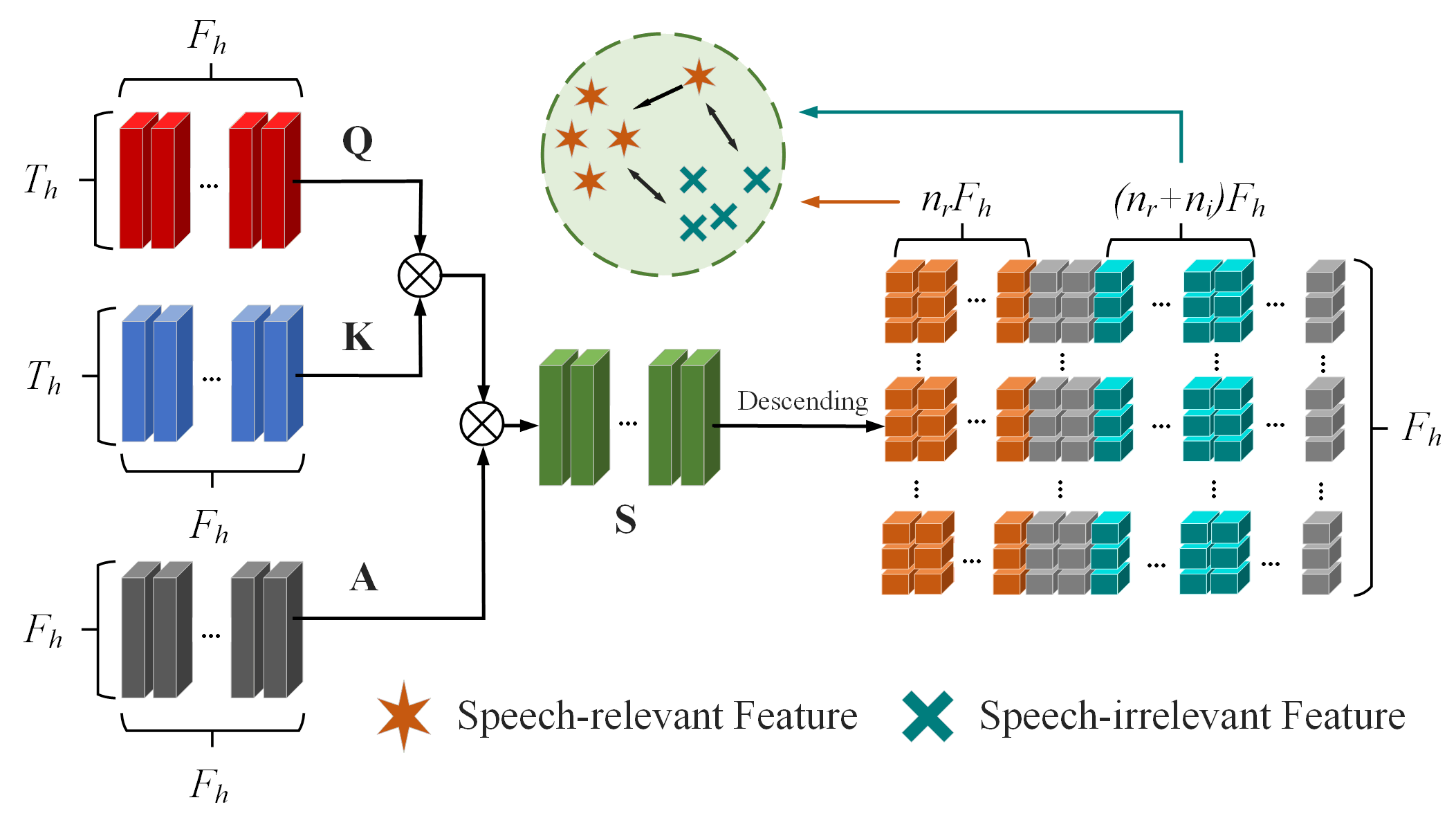}
  \caption{The demonstration of contrastive learning scheme for the proposed contrastive attention module. We define the top $n_rF_h$ features as speech-relevant features and $n_rF_h$ features starting from $n_iF_h$ as speech-irrelevant features for each ordered sequence.}
  \label{fig:4}
\end{figure}

To alleviate the problem, we develop self-attention by applying contrastive learning, called contrastive attention, as shown in Figure~\ref{fig:3}. In particular, we provide a trainable amplification factor $\textbf{A}$ multiplying with $\textbf{Q}^{\top}\textbf{K}$ to enforce the self-attention to give higher aggregation weight on speech-relevant features. As a consequence, the goal of trainable amplification factor $\textbf{A}$ is to increase the gap between relevant and irrelevant features, which is achieved by inserting a loss function based on contrastive learning. As shown in Figure~\ref{fig:4}, the loss for training $\textbf{A}$ can be formulated as:
\begin{equation}
    \textbf{S}=\textbf{A} (\textbf{Q}^{\top}\textbf{K}), \textbf{A}\in \mathbb{R}^{C\times F_h \times F_h}, \{\textbf{Q}, \textbf{K}\}\in \mathbb{R}^{C\times T_h \times F_h}, \label{eq:1}
\end{equation}
\begin{equation}
    \textbf{S}'_{i} = \text{sort}(\textbf{S}_{i}, \text{Descending}), \quad \textbf{S}'_{i}\in \textbf{S}', \quad \textbf{S}_{i}\in \textbf{S},
\end{equation}
\begin{equation}
    \mathcal{L}_{ca} = \frac{1}{F_h}\sum_{i=1}^{F_h}-\log \frac{\sum_{j=1}^{n_rF_h}\text{exp}(\textbf{S}'_{i,j})}{\sum_{j=n_iF_h}^{(n_r+n_i)F_h}\text{exp}(\textbf{S}'_{i,j})} + r,
\end{equation}
where $C$, $T_h$, and $F_h$ indicate the index of the channel, time frame, and frequency bin for the hidden input feature, respectively. $n_r$ and $n_i$ represent the percentage of relevant and irrelevant features in the feature map, and $n_i$ is the start index percentage for irrelevant features in the feature map, respectively. $r$ is a regularization constant. In addition, $\textbf{S}_{i,j}$ measures the relevance between $\textbf{Q}$ and $\textbf{K}$ in self-attention module. $\textbf{S}'_i$ and $\textbf{S}_i$ stand for $i^{th}$ row of $\textbf{S}'$ and $\textbf{S}$, while $\textbf{S}'_i$ is descending sort result of $\textbf{S}_i$.

\begin{figure}[t]
  \centering
  \includegraphics[width=\linewidth]{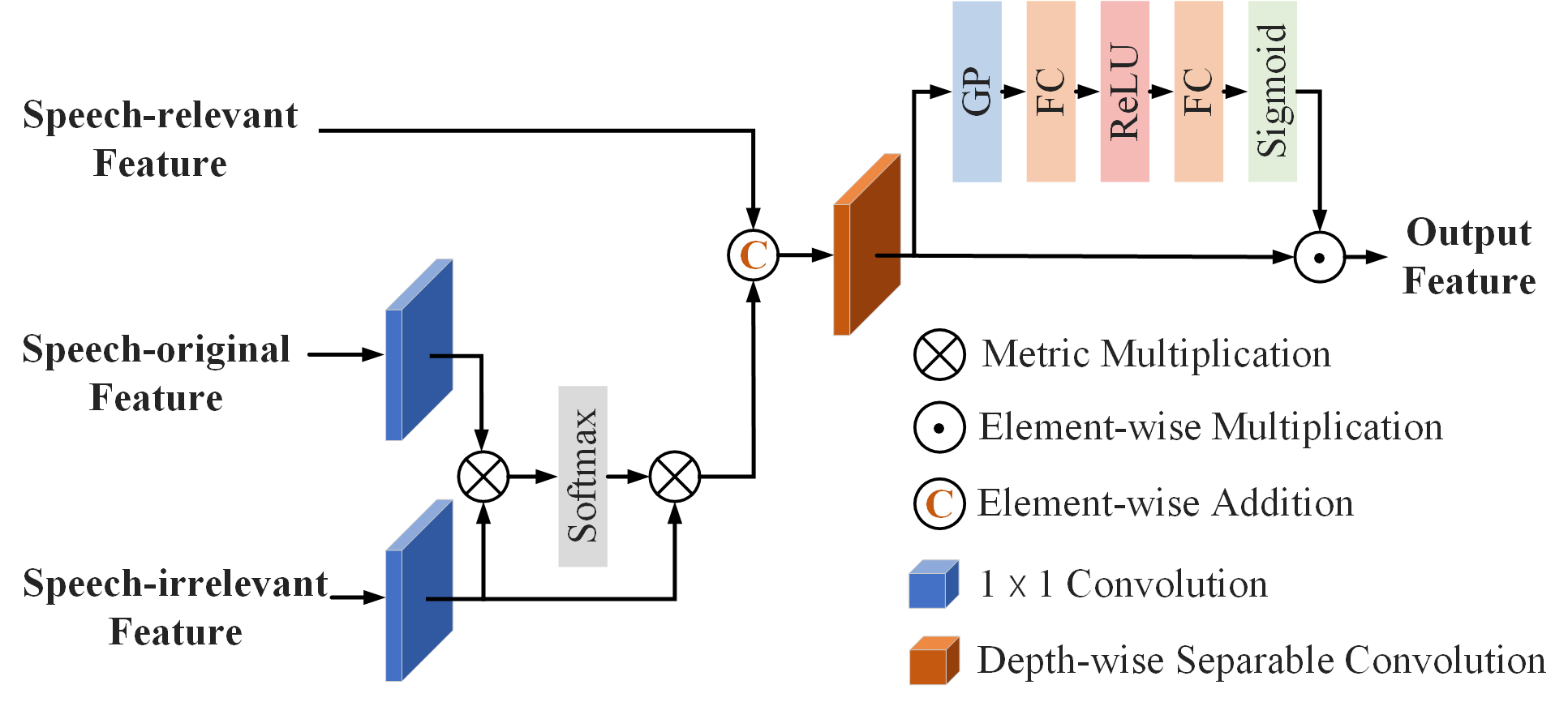}
  \caption{The proposed interactive attention module.}
  \label{fig:5}
\end{figure}

As shown in Figure~\ref{fig:2}(c), the contrastive attention module combined with self-attention and contrastive learning loss effectively filters out more noise components and keeps more speech-relevant features than self-attention. However, the noise components residual and speech-relevant features damage are still hard to avoid when using the contrastive attention module only. To tackle this problem, as shown in Figure~\ref{fig:3}, we extract the speech-relevant features via relevant attention mask, $\textbf{M}_r$, obtained by $\textbf{A} (\textbf{Q}^{\top}\textbf{K})$ applying softmax operation, and also extract the speech-irrelevant features by using irrelevant attention mask, $\textbf{M}_i$, obtained by $\textbf{1}-\textbf{M}_r$. Then, we design an interactive attention module to capture the correlation between speech relevant and irrelevant features for superior performance. As shown in Figure~\ref{fig:5}, the proposed interactive attention module first adopts the self-attention operation to capture the correlation between speech-irrelevant features and speech-original features globally, whose output is then concatenated with speech-relevant features. Finally, the proposed interactive attention module incorporates depth-wise separable convolutions \cite{ref25} and the channel attention layer \cite{ref26} to discriminatively aggregate concatenated features in spatial and channel dimensions \cite{ref27}.

\renewcommand{\arraystretch}{1.05}
\begin{table*}[t]
\centering
\begin{tabular}{cccccccc}
\hline
\multirow{2}{*}{CASE} & \multicolumn{2}{c}{Contrastive Attention} & \multicolumn{2}{c}{Interactive Attetion} & \multirow{2}{*}{PESQ} & \multirow{2}{*}{STOI (\%)} & \multirow{2}{*}{Param. (M)} \\ \cline{2-5}
                      & AF                  & CL                  & DSC                 & CA                 &                       &                            &                                 \\ \hline
1                     & $\times$                   & $\times$                   & $\times$                   & $\times$                  & 2.88                  & 86.47                      & 2.36                            \\
2                     & $\checkmark$                   & $\times$                   & $\times$                   & $\times$                  & 2.93                  & 89.62                      & 2.36                            \\
3                     & $\checkmark$                   & $\checkmark$                   & $\times$                   & $\times$                  & 3.02                  & 91.26                      & 2.36                            \\
4                     & $\checkmark$                   & $\checkmark$                   & $\checkmark$                   & $\times$                  & 3.07                  & 92.69                      & 2.40                            \\
5                     & $\checkmark$                   & $\checkmark$                   & $\checkmark$                   & $\checkmark$                  & 3.10                  & 93.04                      & 2.45                            \\
6                     & $\checkmark$                   & $\times$                   & $\checkmark$                   & $\checkmark$                  & 3.05                  & 91.83                      & 2.45                            \\ 
7                     & $\times$                   & $\times$                   & $\checkmark$                   & $\checkmark$                  & 3.01                  & 90.81                      & 2.45                            \\ \hline
\end{tabular}
\caption{Ablation experiments to study the effectiveness of the proposed CM, contrastive attention, and interactive attention. Note that the ``AF'', `CL'', ``DSC'', and ``CA'' denote the amplification factor, contrastive loss, depth-wise separable convolution, and channel attention, respectively.}
\label{tab:1}
\end{table*}

\subsection{Contrastive Regularization (CR)}

CR follows the principle of contrastive learning that aims to pull the prediction closer to the positive sample while pushing the prediction further away from the negative samples in the representation space. Thus, the CR is proposed to generate better-enhanced speech. 

In this study, we take the group of noisy speech raw waveform $\textbf{y}$ and its enhanced version $\hat{\textbf{s}}$ estimated by CMCR-Net $f(\cdot)$ as the ``negative pair'', while the group of clean speech raw waveform $\textbf{s}$ and estimated speech $\hat{\textbf{s}}$ is treated as the ``positive pair''. Consequently, enhanced speech, clean speech, and noisy speech are treated as anchor, positive sample, and negative sample, respectively. 

In addition, we adopt the pre-trained model as a feature encoder (we evaluate the system performance when adopting different selected pre-trained models in Sec 4) to obtain intermediate features for mapping ``positive'' and ``negative'' pairs in the representation space. Therefore, the CR can be formulated as
\begin{equation}
\begin{aligned}
     \mathcal{L}_{cr} = \frac{L(E(\textbf{s}), E(f(\textbf{y},\theta))))}{L(E(\textbf{y}),E(f(\textbf{y},\theta))))},
\end{aligned}
\end{equation}
where $\textbf{y}$, $\textbf{s}$, and $f(\textbf{y},\theta))$ represent noisy speech waveform, target speech waveform, and estimated speech waveform, respectively, which are under the same latent feature space, which plays a role of opposing forces pulling the enhanced speech $f(\textbf{y},\theta))$ to its clean version $\textbf{s}$ while pushing the enhanced speech $f(\textbf{y},\theta))$ away from the noisy speech $\textbf{y}$. CR improves the performance for SE without introducing additional computation/parameters during the testing phase since they can be removed for inference.

\subsection{Loss Function}
Since the proposed CMCR-Net consists of CM and CR, each containing a loss function. Consequently, three loss functions are used for network training,
\begin{equation}
    \mathcal{L}_{total} = \mathcal{L}_{mse} + \alpha \mathcal{L}_{ca} + \beta\mathcal{L}_{cl},
\end{equation}
where the $\mathcal{L}_{mse}$ calculates the mean-squared error between $\textbf{s}$ and $f(\textbf{y},\theta))$, and $\alpha$ and $\beta$ are scaling coefficients for making these loss values on the same scale.

\section{Experiments}

\subsection{Datasets}
\noindent \textbf{VoiceBank + DEMAND:} \quad This is an open dataset created by \cite{ref28}. Clean speech is collected from Voice Bank \cite{ref29} with 40 male speakers and 40 female speakers from the total of 84 speakers (42 male and 42 female), and each speaker pronounced around 4000 utterances. Afterward, we mix 32,000 utterances with noise from DEMAND \cite{ref30} in $\{-5, -4, -3, -2, -1, 0, 5\}$ dB signal-to-noise ratio (SNR) levels. In addition, we set aside 200 clean utterances from the training set to create a validation set. Test data are generated by mixing 200 utterances selected from the remained 4 untrained speakers (50 each) with noise from DEMAND in three challenge categories: ``Babble'', ``Cafeteria'', and ``Factory'' at three SNR levels $\{-5, 0, 5\}$ dB.

\noindent \textbf{AVSpeech + AudioSet:} \quad This large dataset is used for evaluating the performance of CMCR-Net. The clean dataset AVSpeech is collected from Youtube, and the noisy speech is a mixture of the above clean speech segments with AudioSet \cite{ref31}. In our experiment, 10,000 segments are randomly sampled from the AVSpeech dataset for the training set and 500 segments for the validation dataset. Because of the vast energy distribution in both datasets, the created noisy speech dataset has a wide range of SNR.

\subsection{Implementation}
CRCM-Net is implemented in Pytorch. The dimension of feature maps of encoder are $\{2, 16, 32, 64, 128\}$, and the kernel size of convolutional layers is $2\times 3$ (i.e., $\textit{time} \times \textit{frequency}$). All audios are resampled to 16 kHz. STFT is calculated using the Hann window, whose window length is 25 ms. The hop length is 256 and the FFT size is 512.

The CRCM-Net is trained using Adam optimizer. The initial learning rate and batch size are set to 0.0001 and 16, respectively. Within a minibatch, all training samples are zero-padded to have the same number of time steps as the longest sample. We empirically set the total number of epochs to 50.

Three metrics are used to evaluate CRCM-Net and the state-of-the-art competitors, \textbf{PESQ}, perceptual evaluation of speech quality (from $-0.5 $ to $4.5$), \textbf{STOI}, short-time objective intelligibility measure (from $0$ to $100(\%)$), \textbf{SSNR}, segmental SNR.

\subsection{Ablation Study}
In the ablation study, all the networks are trained with the same random seed. In this section, we perform the ablation studies on the impact of the collaboration module, the selection of feature encoder for contrastive regularization, and the impact of contrastive regularization, and  STOI and PESQ are used on the Voice Bank + DEMAND test set as the evaluation metric.

\subsubsection{Impact of the Collaborate Module (CM)}
To demonstrate the effectiveness of the proposed CM, we construct a baseline model by progressively adding our contrastive attention module and interactive attention module. As shown in Table~\ref{tab:1}, 6 models are compared, \textbf{Case 1}: we replace the contrastive attention module with the self-attention module and remove the interactive attention module, \textbf{Case 2}: we remove the contrastive loss in the contrastive attention module and remove the interactive attention module, \textbf{Case 3}: we remove the interactive attention module, \textbf{Case 4}: we remove the channel attention in the interactive attention module, \textbf{Case 5}: we keep all components of CM, \textbf{Case 6}: we remove the contrastive loss in the contrastive attention module and keep the interactive attention module, and \textbf{Case 7}: we replace the contrastive attention module with the self-attention module and keep the interactive attention module.

\begin{table}[]
\centering
\resizebox{0.36\textwidth}{!}{
\begin{tabular}{lllll}
\hline
\multicolumn{5}{c}{PESQ}                                                                                                   \\ \hline
\multicolumn{1}{l|}{\diagbox{n1}{n2}} & 8     & 16     & 32    & 64   \\ \hline
\multicolumn{1}{l|}{4}                                                  & 3.04  & 3.09  & 3.05  & 3.04\\
\multicolumn{1}{l|}{8}                                                  & 3.07  & 3.10  & 3.06  & 3.04 \\
\multicolumn{1}{l|}{10}                                                  & 3.05  & 3.08  & 3.06  & 3.03 \\ \hline
\multicolumn{5}{c}{STOI (\%)}                                                                                               \\ \hline
\multicolumn{1}{l|}{\diagbox{n1}{n2}} & 8     & 16     & 32    & 64                        \\ \hline
\multicolumn{1}{l|}{4}                                                  & 91.89 & 92.87 & 92.96 & 92.43                     \\
\multicolumn{1}{l|}{8}                                                  & 92.92 & 93.04 & 93.15 & 93.01                     \\
\multicolumn{1}{l|}{10}                                                  & 92.74 & 92.76 & 92.88 & 92.26                     \\ \hline
\end{tabular}}
\caption{The ablation experiment to explore the effects of $n_1 (\%)$ and $n_2 (\%)$ on contrastive loss.}
\label{tab:2}
\end{table}

\textbf{The effectiveness of contrastive attention module.} To verify the effectiveness of the contrastive attention module, we first compare \textbf{Case 1} with \textbf{Case 3}. Our contrastive attention module brings 0.14 PESQ and $4.97\%$ STOI gains over self-attention, which demonstrates the effectiveness of the contrastive attention module. Furthermore, by solely adding the components such as adding the amplification factor \textbf{A} in Eq~\ref{eq:1} and contrastive learning, results further improve around 0.05 improvement in PESQ and $3.15\%$ improvement in STOI by comparing \textbf{Case 1} with \textbf{Case 2}. Compare \textbf{Case 2} with \textbf{Case 3}, using contrastive learning achieves around 0.09 improvement in PESQ and $1.64\%$ improvement in STOI. In contrastive loss, the $n_1$ determines the percentage of relevant and irrelevant features participating in contrastive learning, and $n_2$ indicates the start index percentage of irrelevant features in the feature map. Results shown in Table~\ref{tab:2} show that setting $n_1$ to $8\%$ and setting $n_2$ to $16\%$ leads best performance.

\textbf{The impact of speech-relevant and speech-irrelevant features interaction strategy.} The models from \textbf{Case 1} to \textbf{Case 3} uses the speech-relevant features only. To verify the significance of speech-irrelevant features, we set 3 groups for comparison, which are \textbf{Case 1} and \textbf{Case 7}, \textbf{Case 2} and \textbf{Case 6}, and \textbf{Case 3} and \textbf{Case 5}. According to Table~\ref{tab:1}, we observe that \textbf{Case 5}, \textbf{Case 6}, and \textbf{Case 7} are outperform the \textbf{Case 3}, \textbf{Case 2}, and \textbf{Case 1}, respectively. The comparison results based on these three groups fully demonstrate the validity of using speech-irrelevant features for performance improvement.

\renewcommand{\arraystretch}{0.95}
\begin{table}[]
\centering
\resizebox{0.47\textwidth}{!}{
\begin{tabular}{cccc}
\hline
Model                  & PESQ & STOI (\%) & Param. (M) \\ \hline
Interactive Attention  & 3.10 & 93.04     & 2.45       \\
Concatenation Operation & 2.96 & 90.29     & 2.38       \\
Self-attention         & 3.02 & 91.33     & 2.57       \\
SKNet                  & 2.98 & 91.05     & 2.42       \\ \hline
\end{tabular}}
\caption{Ablation study on interactive attention module.}
\label{tab:3}
\end{table}

\textbf{The impact of interactive attention module.} We conduct the ablation study to validate the contribution of interactive attention including depth-wise separable convolution (DSC) and channel attention. According to the comparison between \textbf{Case 3} and \textbf{Case 4}, we observe that using the depth-wise separable convolution allows increasing the channel depth with approximately the same parameters and enhance the representation capacity. The comparison between \textbf{Case 5} and \textbf{Case 6} demonstrates that the effectiveness of channel attention for enhancing the representation capacity. To further verify the effectiveness of interactive attention module, we evaluate the cases when we separately replace the interactive attention module with a simple concatenation, a self-attention, and a SKNet \cite{ref32}. Table~\ref{tab:3} effectively indicates the significance of the proposed interactive attention module.

\begin{figure}[t]
\centering
\subfigure[Noisy input]{
\label{Fig.sub.1}
\includegraphics[width=0.195\textwidth]{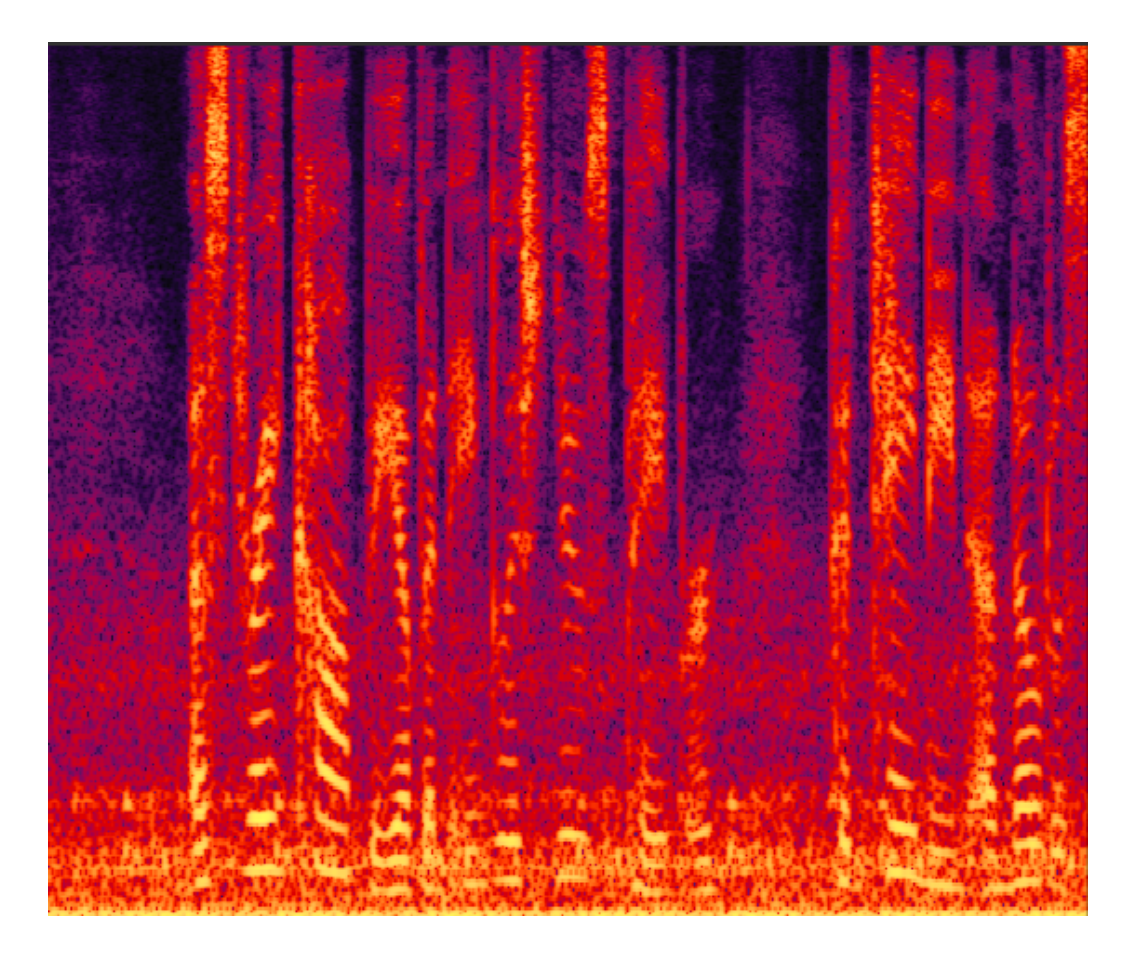}
}
\subfigure[Only MSE loss]{
\label{Fig.sub.2}
\includegraphics[width=0.195\textwidth]{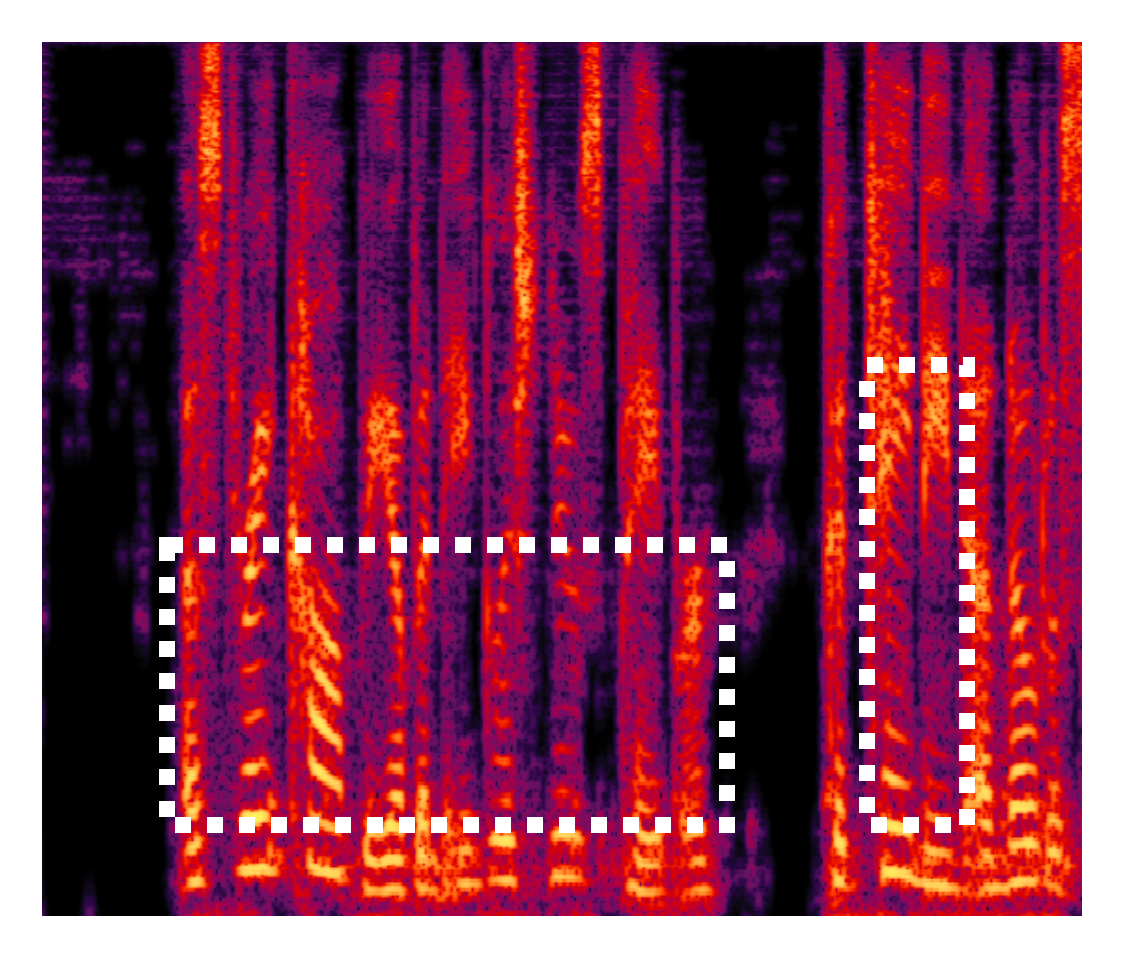}
\centering}
\qquad

\subfigure[Our MSE + CR]{
\label{Fig.sub.3}
\includegraphics[width=0.193\textwidth]{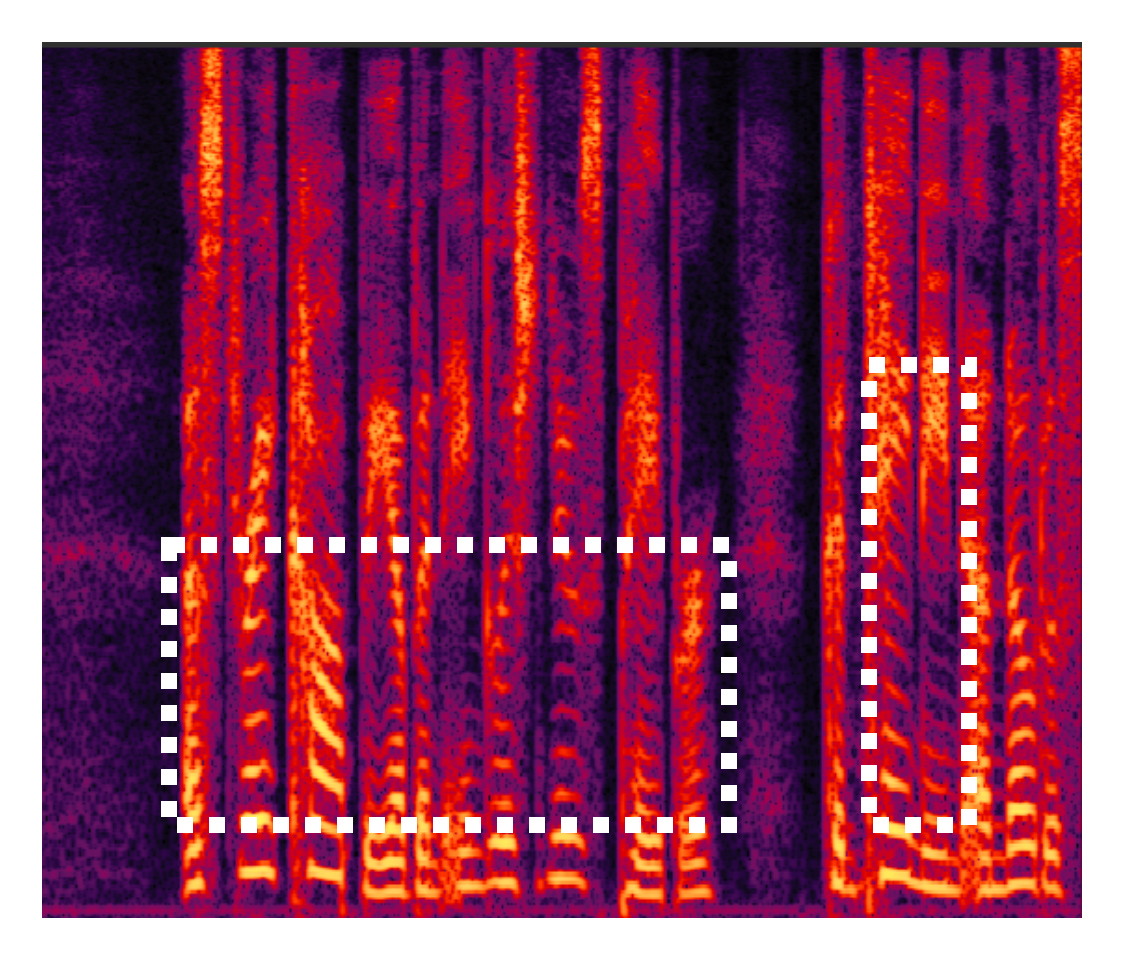}
\centering}
\subfigure[Ground-truth]{
\label{Fig.sub.4}
\includegraphics[width=0.198\textwidth]{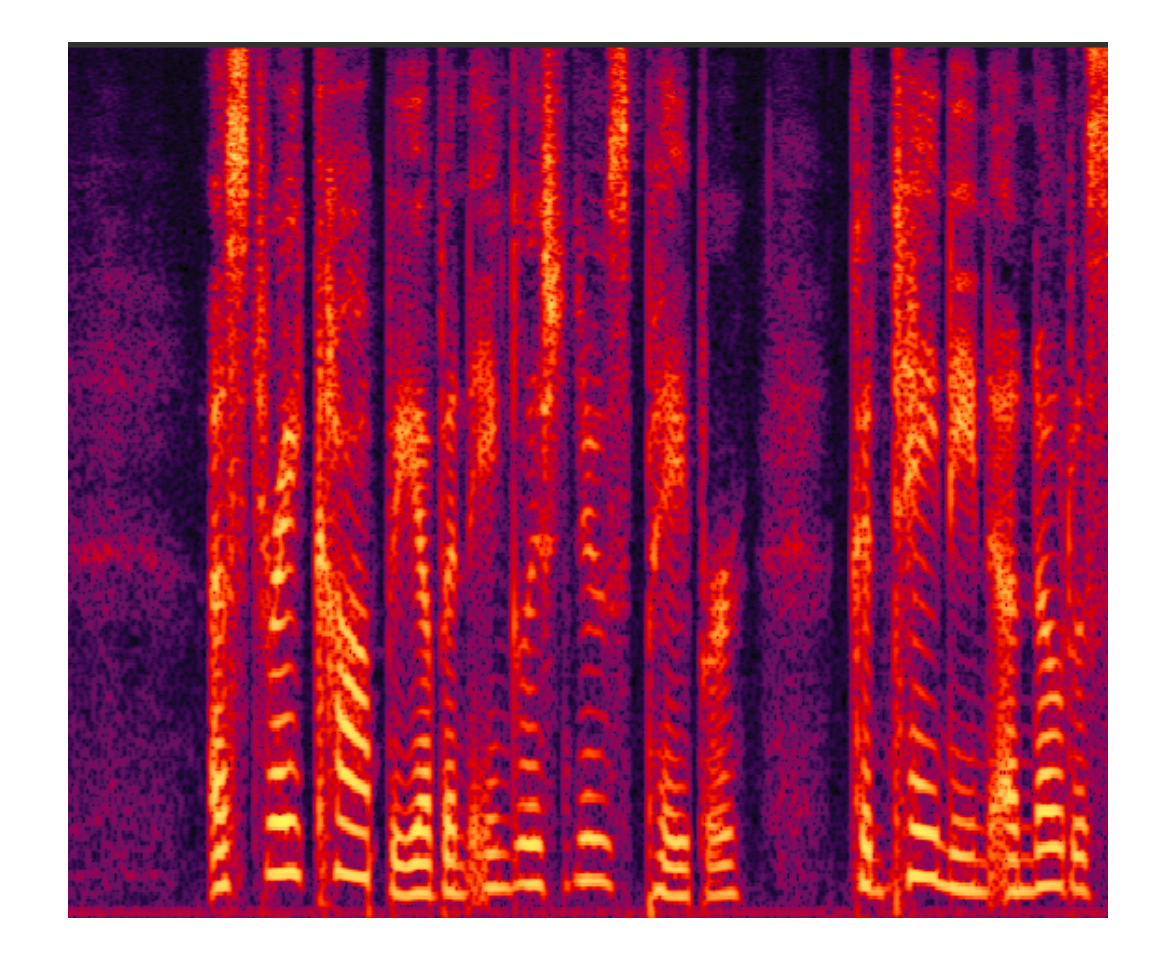}
\centering}
\caption{Comparison between only positive-orient supervision and joint negative- and positive-orient supervision.}
\label{fig:6}
\end{figure}

\renewcommand{\arraystretch}{0.95}
\begin{table}[]
\centering
\resizebox{0.47\textwidth}{!}{
\begin{tabular}{lccc}
\hline
Model             & Trained Dataset & PESQ & STOI (\%) \\ \hline
None              & -               & 2.82 & 89.86     \\ \hline
vq-wav2vec        & Clean           & 2.96 & 91.79     \\
wav2vec2          & Noisy           & 3.01 & 92.34     \\
HuBERT            & Noisy           & 3.05 & 92.96     \\
WavLM             & Clean           & 3.08 & 93.12     \\ \hline
UniSpeech-SAT     & Noisy           & 3.10 & 93.04     \\
\quad-w/o Noisy Sample & -               & 3.01 & 92.80     \\
\quad-w/o Clean Sample & -               & 2.92 & 91.39     \\ \hline
\end{tabular}}
\caption{Evaluating 5 pre-training models for feature encoder.}
\label{tab:4}
\end{table}

\subsubsection{Impact of Contrastive Regularization (CR)}
To verify the effectiveness of CR, we firstly select 5 pre-training models, i.e., vq-wav2vec \cite{ref34}, wave2vec2 \cite{ref35}, HuBERT \cite{ref36}, WavLM \cite{ref37}, and UniSpeech-SAT \cite{ref33} as feature encoder in CR for comparison, and the results are shown in Table~\ref{tab:4}. Note that ``None'' denotes that the system does not apply CR and UniSpeech-SAT, WavLM, and wav2vec2 have seen noisy speech while other models are pre-trained on the clean speech from audiobooks. According to the table, we observe that the CR significantly improves the system performance whatever the pre-training model we use. Furthermore, UniSpeech-SAT yields the best PESQ score while WavLM yields the best STOI score. Interestingly, the results are suitable for the findings in \cite{ref38}, and both noisy and clean samples in CR do improve the system performance. Therefore, for good accuracy, we use UniSpeech-SAT as a feature encoder for the final CM to do the comparison. The visualization of the enhanced speech spectrogram by SE model with and without CR is shown in Figure~\ref{fig:6}.

\renewcommand{\arraystretch}{0.95}
\begin{table}[]
\centering
\resizebox{0.37\textwidth}{!}{
\begin{tabular}{ccc}
\hline
Model   & STOI(\%)     & PESQ        \\ \hline
DCCRN  & 87.71 (${\color{red}\uparrow}$1.60) & 2.73 (${\color{red}\uparrow}$0.06) \\
TFT-Net    & 89.58 (${\color{red}\uparrow}$1.43) & 2.84 (${\color{red}\uparrow}$0.07) \\
U-Former & 91.32 (${\color{red}\uparrow}$1.28) & 3.02 (${\color{red}\uparrow}$0.04) \\ \hline
\end{tabular}}
\caption{Impact of applying CR into other SE models.}
\label{tab:5}
\end{table}

To evaluate the university of the proposed CR, we add CR into three other SE models, i.e., DCCRN \cite{ref23}, TFT-Net \cite{ref7}, and U-Former \cite{ref39}. As presented in Table~\ref{tab:5}, CR can further improve the performance of these SE models. In other words, our CR is model agnostic to train the SE networks effectively. Furthermore, our CR cannot increase the additional parameters for inference, since it can be directly removed for testing.

\renewcommand{\arraystretch}{0.95}
\begin{table}[]
\centering
\resizebox{0.48\textwidth}{!}{
\begin{tabular}{lcccc}
\hline
Model                               & PESQ & STOI (\%) & Param. & PTs(s) \\ \hline
Unproessed                          & 1.97 & 71.26     & -      & -      \\ \hline
DCCRN                               & 2.67 & 86.61     & 3.67   & 0.43   \\
PHASEN                              & 2.89 & 89.79     & 6.41   & 0.64   \\
SA-TCN                              & 2.95 & 90.13     & 5.83   & 0.71   \\
U-Former                            & 2.98 & 90.04     & 2.78   & 0.68   \\ \hline
CMCR-Net                            & 3.10 & 93.04     & 2.45   & 0.42   \\
\multicolumn{1}{c}{CMCR-Net-w/o CR} & 3.02 & 91.26     & 2.45   & 0.42   \\ \hline
\end{tabular}}
\caption{System comparison on Voice Bank + DEMAND.}
\label{tab:6}
\end{table}

\subsection{Comparison with the State-of-the-Art}
\noindent \textbf{Voice Bank + BEMAND:} We train the proposed CMCR-Net on synthetic and commonly-used dataset Voice Bank + DEMAND. We fairly compare out CMCR-Net with 4 other methods, DCCRN, PHASEN \cite{ref40}, SA-TCN \cite{ref41}, and U-Former. Table~\ref{tab:3} shows the comparison results in terms of STOI and PESQ. It can be seen that the proposed CMCR-Net achieves the best performance on PESQ and STOI. We also report the processing time of these models. Note that the processing time of these models is calculated with an input noisy speech of 16,000 samples (1 second). According to Table~\ref{tab:6}, we observe that CMCR-Net yields the lowest model size and processing time. 

\begin{figure}[]
\centering
\subfigure[STOI score ($\%$)]{
\label{Fig7.sub.1}
\includegraphics[width=0.223\textwidth]{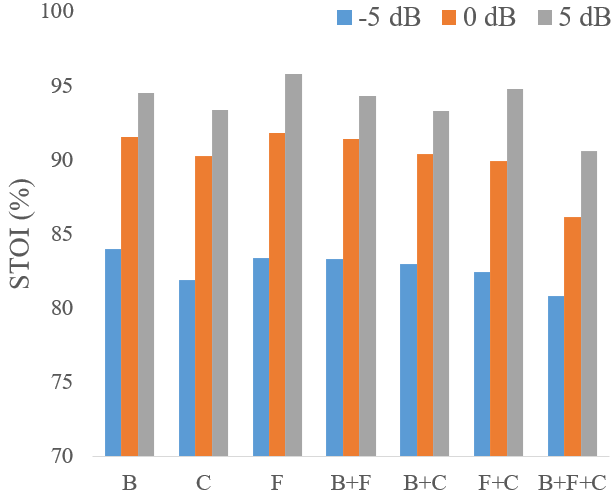}}
\subfigure[PESQ score]{
\label{Fig7.sub.2}
\includegraphics[width=0.223\textwidth]{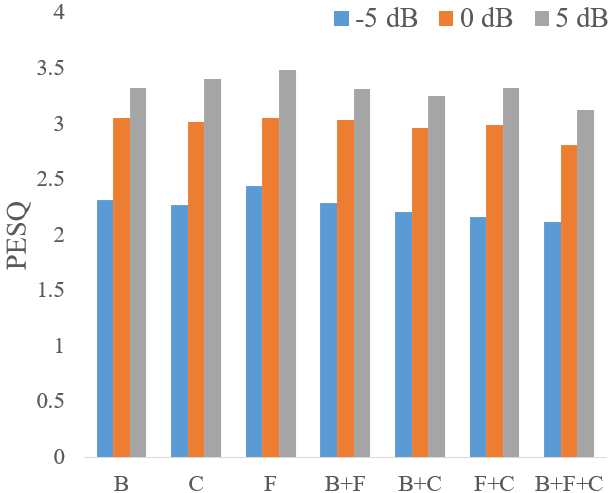}}
\caption{Comparison in STOI and PESQ average for cases with multiple noise types, in which B, F, and C denote babble, factory, and cafeteria noisy types, respectively.}
\label{fig:7}
\end{figure}

\renewcommand{\arraystretch}{0.95}
\begin{table}[t]
\centering
\resizebox{0.48\textwidth}{!}{
\begin{tabular}{cccc}
\hline
Metrics & SSNR (dB)          & PESQ          & STOI (\%)      \\ \hline
Attention-U-Net (2019)   & 7.34           & 2.34          & 83.94          \\
DCCRN (2020)    & 9.22           & 2.88          & 86.50          \\
PHASEN (2020) & 10.89          & 3.19          & 88.41          \\
TFT-Net (2021) & 10.62          & 3.18          & 89.32          \\
U-Former (2022)  & 10.34          & 3.16          & 88.87          \\
SA-TCN (2021)  & 11.06          & 3.20          & 88.12          \\ \hline
CMCR-Net & \textbf{11.64} & \textbf{3.26} & \textbf{90.53} \\ \hline
\end{tabular}}
\caption{System comparison on AVspeech + Audioset}
\label{tab:7}
\end{table}

To demonstrate the effectiveness and flexibility of the CMCR-Net, we further explore and apply CMCR-Net to circumstance when multiple noise types are included in noisy background. We randomly select 150 clean utterances from test set and 3 challenging noises, babble, factory, and cafeteria, which are explained in Section 4.1, to produce the multiple types of noises with clean utterances at -5 dB, 0 dB, and 5 dB SNR levels.  Figure~\ref{fig:7} present the PESQ and STOI results over three SNR levels on multiple target noise types, in which B, F, and C denote babble, factory, and cafeteria noisy types, respectively. From Figure~\ref{fig:7}, we observe that the results using 1, 2, and 3 noise types are comparable, although the STOI and PESQ scores are decreasing when the number of noise types is increase.

\noindent \textbf{AVSpeech + AudioSet:} \quad On this large dataset, we compare the proposed CMCR-Net with several state-of-the-art methods. Attention-U-Net \cite{ref42}, DCCRN, PHASEN, TFT-Net, SA-TCN, and U-Former increased results in Table~\ref{tab:7} demonstrate that CRCM-Net outperforms these state-of-the-art methods and also indicate that CRCM-Net can be generalized to various speakers and various kinds of noisy environments. It suggests that CMCR-Net is readily applicable to complicated real-world environment. 

\section{Conclusion}

In this paper, we propose a novel CMCR-Net for monaural speech enhancement, which consists of CR and CM under encoder-decoder architecture. CM module consists of contrastive attention and interactive attention, which establish the correlation between the speech relevant and irrelevant information at the global level and local-level in a learnable and self-adaptive manner. CR is build upon contrastive learning to ensure that the enhanced speech is pulled to closer to the clean speech while pushing to far away from the noisy speech in representation space.  Experimental results show that the proposed method is capable of superior performance in SE with attractive computation costs.

\vfill\pagebreak
\bibliographystyle{named}
\bibliography{ijcai22}

\begin{thebibliography}{}

\bibitem[\protect\citeauthoryear{Baevski \bgroup \em et al.\egroup
  }{2019}]{ref34}
Alexei Baevski, Steffen Schneider, and Michael Auli.
\newblock vq-wav2vec: Self-supervised learning of discrete speech
  representations.
\newblock In {\em International Conference on Learning Representations}, 2019.

\bibitem[\protect\citeauthoryear{Baevski \bgroup \em et al.\egroup
  }{2020}]{ref35}
Alexei Baevski, Yuhao Zhou, Abdelrahman Mohamed, and Michael Auli.
\newblock wav2vec 2.0: A framework for self-supervised learning of speech
  representations.
\newblock {\em Advances in Neural Information Processing Systems},
  33:12449--12460, 2020.

\bibitem[\protect\citeauthoryear{Boll}{1979}]{ref11}
Steven Boll.
\newblock Suppression of acoustic noise in speech using spectral subtraction.
\newblock {\em IEEE Transactions on acoustics, speech, and signal processing},
  27(2):113--120, 1979.

\bibitem[\protect\citeauthoryear{Chen \bgroup \em et al.\egroup }{2020}]{ref19}
Ting Chen, Simon Kornblith, Mohammad Norouzi, and Geoffrey Hinton.
\newblock A simple framework for contrastive learning of visual
  representations.
\newblock In {\em International conference on machine learning}, pages
  1597--1607. PMLR, 2020.

\bibitem[\protect\citeauthoryear{Chen \bgroup \em et al.\egroup
  }{2022a}]{ref37}
Sanyuan Chen, Chengyi Wang, Zhengyang Chen, Yu~Wu, Shujie Liu, Zhuo Chen, Jinyu
  Li, Naoyuki Kanda, Takuya Yoshioka, Xiong Xiao, et~al.
\newblock Wavlm: Large-scale self-supervised pre-training for full stack speech
  processing.
\newblock {\em IEEE Journal of Selected Topics in Signal Processing},
  16(6):1505--1518, 2022.

\bibitem[\protect\citeauthoryear{Chen \bgroup \em et al.\egroup
  }{2022b}]{ref33}
Sanyuan Chen, Yu~Wu, Chengyi Wang, Zhengyang Chen, Zhuo Chen, Shujie Liu, Jian
  Wu, Yao Qian, Furu Wei, Jinyu Li, et~al.
\newblock Unispeech-sat: Universal speech representation learning with speaker
  aware pre-training.
\newblock In {\em ICASSP 2022-2022 IEEE International Conference on Acoustics,
  Speech and Signal Processing (ICASSP)}, pages 6152--6156. IEEE, 2022.

\bibitem[\protect\citeauthoryear{Chollet}{2017}]{ref25}
Fran{\c{c}}ois Chollet.
\newblock Xception: Deep learning with depthwise separable convolutions.
\newblock In {\em Proceedings of the IEEE conference on computer vision and
  pattern recognition}, pages 1251--1258, 2017.

\bibitem[\protect\citeauthoryear{Cohen and Berdugo}{2002}]{ref10}
Israel Cohen and Baruch Berdugo.
\newblock Noise estimation by minima controlled recursive averaging for robust
  speech enhancement.
\newblock {\em IEEE signal processing letters}, 9(1):12--15, 2002.

\bibitem[\protect\citeauthoryear{Ephrat \bgroup \em et al.\egroup
  }{2018}]{ref13}
Ariel Ephrat, Inbar Mosseri, Oran Lang, Tali Dekel, Kevin Wilson, Avinatan
  Hassidim, William~T Freeman, and Michael Rubinstein.
\newblock Looking to listen at the cocktail party: a speaker-independent
  audio-visual model for speech separation.
\newblock {\em ACM Transactions on Graphics (TOG)}, 37(4):1--11, 2018.

\bibitem[\protect\citeauthoryear{Fu \bgroup \em et al.\egroup }{2017}]{ref16}
Szu-Wei Fu, Yu~Tsao, Xugang Lu, and Hisashi Kawai.
\newblock Raw waveform-based speech enhancement by fully convolutional
  networks.
\newblock In {\em 2017 Asia-Pacific Signal and Information Processing
  Association Annual Summit and Conference (APSIPA ASC)}, pages 006--012. IEEE,
  2017.

\bibitem[\protect\citeauthoryear{Gemmeke \bgroup \em et al.\egroup
  }{2017}]{ref31}
Jort~F Gemmeke, Daniel~PW Ellis, Dylan Freedman, Aren Jansen, Wade Lawrence,
  R~Channing Moore, Manoj Plakal, and Marvin Ritter.
\newblock Audio set: An ontology and human-labeled dataset for audio events.
\newblock In {\em 2017 IEEE international conference on acoustics, speech and
  signal processing (ICASSP)}, pages 776--780. IEEE, 2017.

\bibitem[\protect\citeauthoryear{Giri \bgroup \em et al.\egroup }{2019}]{ref42}
Ritwik Giri, Umut Isik, and Arvindh Krishnaswamy.
\newblock Attention wave-u-net for speech enhancement.
\newblock In {\em 2019 IEEE Workshop on Applications of Signal Processing to
  Audio and Acoustics (WASPAA)}, pages 249--253. IEEE, 2019.

\bibitem[\protect\citeauthoryear{He \bgroup \em et al.\egroup }{2020}]{ref20}
Kaiming He, Haoqi Fan, Yuxin Wu, Saining Xie, and Ross Girshick.
\newblock Momentum contrast for unsupervised visual representation learning.
\newblock In {\em Proceedings of the IEEE/CVF conference on computer vision and
  pattern recognition}, pages 9729--9738, 2020.

\bibitem[\protect\citeauthoryear{Hsu \bgroup \em et al.\egroup }{2021}]{ref36}
Wei-Ning Hsu, Benjamin Bolte, Yao-Hung~Hubert Tsai, Kushal Lakhotia, Ruslan
  Salakhutdinov, and Abdelrahman Mohamed.
\newblock Hubert: Self-supervised speech representation learning by masked
  prediction of hidden units.
\newblock {\em IEEE/ACM Transactions on Audio, Speech, and Language
  Processing}, 29:3451--3460, 2021.

\bibitem[\protect\citeauthoryear{Hu \bgroup \em et al.\egroup }{2018}]{ref26}
Jie Hu, Li~Shen, and Gang Sun.
\newblock Squeeze-and-excitation networks.
\newblock In {\em Proceedings of the IEEE conference on computer vision and
  pattern recognition}, pages 7132--7141, 2018.

\bibitem[\protect\citeauthoryear{Hu \bgroup \em et al.\egroup }{2020}]{ref23}
Yanxin Hu, Yun Liu, Shubo Lv, Mengtao Xing, Shimin Zhang, Yihui Fu, Jian Wu,
  Bihong Zhang, and Lei Xie.
\newblock {DCCRN: Deep Complex Convolution Recurrent Network for Phase-Aware
  Speech Enhancement}.
\newblock {\em Proc. Interspeech 2020}, pages 2472--2476, 2020.

\bibitem[\protect\citeauthoryear{Huang \bgroup \em et al.\egroup
  }{2022}]{ref38}
Zili Huang, Shinji Watanabe, Shu-wen Yang, Paola Garc{\'\i}a, and Sanjeev
  Khudanpur.
\newblock Investigating self-supervised learning for speech enhancement and
  separation.
\newblock In {\em ICASSP 2022-2022 IEEE International Conference on Acoustics,
  Speech and Signal Processing (ICASSP)}, pages 6837--6841. IEEE, 2022.

\bibitem[\protect\citeauthoryear{Jiang \bgroup \em et al.\egroup
  }{2022}]{ref27}
Kui Jiang, Zhongyuan Wang, Chen Chen, Zheng Wang, Laizhong Cui, and Chia-Wen
  Lin.
\newblock {Magic ELF}: Image deraining meets association learning and
  transformer.
\newblock In {\em Proceedings of the 30th ACM International Conference on
  Multimedia}, pages 827--836, 2022.

\bibitem[\protect\citeauthoryear{Li \bgroup \em et al.\egroup }{2019}]{ref32}
Xiang Li, Wenhai Wang, Xiaolin Hu, and Jian Yang.
\newblock Selective kernel networks.
\newblock In {\em Proceedings of the IEEE/CVF conference on computer vision and
  pattern recognition}, pages 510--519, 2019.

\bibitem[\protect\citeauthoryear{Lin \bgroup \em et al.\egroup }{2021}]{ref41}
Ju~Lin, Adriaan J de~Lind van Wijngaarden, Kuang-Ching Wang, and Melissa~C
  Smith.
\newblock Speech enhancement using multi-stage self-attentive temporal
  convolutional networks.
\newblock {\em IEEE/ACM Transactions on Audio, Speech, and Language
  Processing}, 29:3440--3450, 2021.

\bibitem[\protect\citeauthoryear{Liu \bgroup \em et al.\egroup }{2021}]{ref12}
Wenzhe Liu, Andong Li, Yuxuan Ke, Chengshi Zheng, and Xiaodong Li.
\newblock Know your enemy, know yourself: A unified two-stage framework for
  speech enhancement.
\newblock In {\em Interspeech}, pages 186--190, 2021.

\bibitem[\protect\citeauthoryear{Ohishi \bgroup \em et al.\egroup
  }{2022}]{ref21}
Yasunori Ohishi, Marc Delcroix, Tsubasa Ochiai, Shoko Araki, Daiki Takeuchi,
  Daisuke Niizumi, Akisato Kimura, Noboru Harada, and Kunio Kashino.
\newblock Conceptbeam: Concept driven target speech extraction.
\newblock In {\em Proceedings of the 30th ACM International Conference on
  Multimedia}, pages 4252--4260, 2022.

\bibitem[\protect\citeauthoryear{Park and Lee}{2017}]{ref17}
Se~Rim Park and Jin~Won Lee.
\newblock A fully convolutional neural network for speech enhancement.
\newblock {\em Proc. Interspeech 2017}, pages 1993--1997, 2017.

\bibitem[\protect\citeauthoryear{Park \bgroup \em et al.\egroup }{2020}]{ref22}
Taesung Park, Alexei~A Efros, Richard Zhang, and Jun-Yan Zhu.
\newblock Contrastive learning for unpaired image-to-image translation.
\newblock In {\em European Conference on Computer Vision}, pages 319--345.
  Springer, 2020.

\bibitem[\protect\citeauthoryear{Peng \bgroup \em et al.\egroup }{2022}]{ref1}
Yifan Peng, Siddharth Dalmia, Ian Lane, and Shinji Watanabe.
\newblock Branchformer: Parallel mlp-attention architectures to capture local
  and global context for speech recognition and understanding.
\newblock In {\em International Conference on Machine Learning}, pages
  17627--17643. PMLR, 2022.

\bibitem[\protect\citeauthoryear{Rao \bgroup \em et al.\egroup }{2019}]{ref2}
Wei Rao, Chenglin Xu, Eng~Siong Chng, and Haizhou Li.
\newblock Target speaker extraction for multi-talker speaker verification.
\newblock {\em Proc. Interspeech 2019}, pages 1273--1277, 2019.

\bibitem[\protect\citeauthoryear{Sell \bgroup \em et al.\egroup }{2018}]{ref3}
Gregory Sell, David Snyder, Alan McCree, Daniel Garcia-Romero, Jes{\'u}s
  Villalba, Matthew Maciejewski, Vimal Manohar, Najim Dehak, Daniel Povey,
  Shinji Watanabe, et~al.
\newblock Diarization is hard: Some experiences and lessons learned for the jhu
  team in the inaugural {DIHARD} challenge.
\newblock In {\em Interspeech}, pages 2808--2812, 2018.

\bibitem[\protect\citeauthoryear{Sermanet \bgroup \em et al.\egroup
  }{2018}]{ref18}
Pierre Sermanet, Corey Lynch, Yevgen Chebotar, Jasmine Hsu, Eric Jang, Stefan
  Schaal, Sergey Levine, and Google Brain.
\newblock Time-contrastive networks: Self-supervised learning from video.
\newblock In {\em 2018 IEEE international conference on robotics and automation
  (ICRA)}, pages 1134--1141. IEEE, 2018.

\bibitem[\protect\citeauthoryear{Tan and Wang}{2019}]{ref24}
Ke~Tan and DeLiang Wang.
\newblock Learning complex spectral mapping with gated convolutional recurrent
  networks for monaural speech enhancement.
\newblock {\em IEEE/ACM Transactions on Audio, Speech, and Language
  Processing}, 28:380--390, 2019.

\bibitem[\protect\citeauthoryear{Tang \bgroup \em et al.\egroup }{2021}]{ref7}
Chuanxin Tang, Chong Luo, Zhiyuan Zhao, Wenxuan Xie, and Wenjun Zeng.
\newblock Joint time-frequency and time domain learning for speech enhancement.
\newblock In {\em Proceedings of the Twenty-Ninth International Conference on
  International Joint Conferences on Artificial Intelligence}, pages
  3816--3822, 2021.

\bibitem[\protect\citeauthoryear{Thiemann \bgroup \em et al.\egroup
  }{2013}]{ref30}
Joachim Thiemann, Nobutaka Ito, and Emmanuel Vincent.
\newblock The diverse environments multi-channel acoustic noise database
  (demand): A database of multichannel environmental noise recordings.
\newblock In {\em Proceedings of Meetings on Acoustics ICA2013}, volume~19,
  page 035081. Acoustical Society of America, 2013.

\bibitem[\protect\citeauthoryear{Valentini-Botinhao \bgroup \em et al.\egroup
  }{2016}]{ref28}
Cassia Valentini-Botinhao, Xin Wang, Shinji Takaki, and Junichi Yamagishi.
\newblock Investigating rnn-based speech enhancement methods for noise-robust
  text-to-speech.
\newblock In {\em SSW}, pages 146--152, 2016.

\bibitem[\protect\citeauthoryear{Veaux \bgroup \em et al.\egroup
  }{2013}]{ref29}
Christophe Veaux, Junichi Yamagishi, and Simon King.
\newblock The voice bank corpus: Design, collection and data analysis of a
  large regional accent speech database.
\newblock In {\em 2013 international conference oriental COCOSDA}, pages 1--4.
  IEEE, 2013.

\bibitem[\protect\citeauthoryear{Xu and Hao}{2022}]{ref39}
Xinmeng Xu and Jianjun Hao.
\newblock {U-Former: Improving Monaural Speech Enhancement with Multi-head Self
  and Cross Attention}.
\newblock In {\em 2022 26th International Conference on Pattern Recognition
  (ICPR)}, pages 663--669, 2022.

\bibitem[\protect\citeauthoryear{Xu \bgroup \em et al.\egroup }{2013}]{ref4}
Yong Xu, Jun Du, Li-Rong Dai, and Chin-Hui Lee.
\newblock An experimental study on speech enhancement based on deep neural
  networks.
\newblock {\em IEEE Signal processing letters}, 21(1):65--68, 2013.

\bibitem[\protect\citeauthoryear{Xu \bgroup \em et al.\egroup }{2021}]{ref5}
Xinmeng Xu, Yang Wang, Dongxiang Xu, Yiyuan Peng, Cong Zhang, Jie Jia, and
  Binbin Chen.
\newblock Multi-stage progressive speech enhancement network.
\newblock In {\em Interspeech}, pages 2691--2695, 2021.

\bibitem[\protect\citeauthoryear{Xu \bgroup \em et al.\egroup }{2022a}]{ref6}
Xinmeng Xu, Rongzhi Gu, and Yuexian Zou.
\newblock Improving dual-microphone speech enhancement by learning
  cross-channel features with multi-head attention.
\newblock In {\em ICASSP 2022-2022 IEEE International Conference on Acoustics,
  Speech and Signal Processing (ICASSP)}, pages 6492--6496. IEEE, 2022.

\bibitem[\protect\citeauthoryear{Xu \bgroup \em et al.\egroup }{2022b}]{ref15}
Xinmeng Xu, Yang Wang, Jie Jia, Binbin Chen, and Jianjun Hao.
\newblock {GLD-Net: Improving Monaural Speech Enhancement by Learning Global
  and Local Dependency Features with GLD Block}.
\newblock In {\em Proc. Interspeech 2022}, pages 966--970, 2022.

\bibitem[\protect\citeauthoryear{Yin \bgroup \em et al.\egroup }{2020}]{ref40}
Dacheng Yin, Chong Luo, Zhiwei Xiong, and Wenjun Zeng.
\newblock Phasen: A phase-and-harmonics-aware speech enhancement network.
\newblock In {\em Proceedings of the AAAI Conference on Artificial
  Intelligence}, volume~34, pages 9458--9465, 2020.

\bibitem[\protect\citeauthoryear{Zhang \bgroup \em et al.\egroup }{2020}]{ref9}
Qiquan Zhang, Aaron Nicolson, Mingjiang Wang, Kuldip~K Paliwal, and Chenxu
  Wang.
\newblock {DeepMMSE}: A deep learning approach to mmse-based noise power
  spectral density estimation.
\newblock {\em IEEE/ACM Transactions on Audio, Speech, and Language
  Processing}, 28:1404--1415, 2020.

\bibitem[\protect\citeauthoryear{Zhao \bgroup \em et al.\egroup }{2022}]{ref14}
Haoran Zhao, Nan Li, Runqiang Han, Lianwu Chen, Xiguang Zheng, Chen Zhang,
  Liang Guo, and Bing Yu.
\newblock A deep hierarchical fusion network for fullband acoustic echo
  cancellation.
\newblock In {\em ICASSP 2022-2022 IEEE International Conference on Acoustics,
  Speech and Signal Processing (ICASSP)}, pages 9112--9116. IEEE, 2022.

\bibitem[\protect\citeauthoryear{Zheng \bgroup \em et al.\egroup }{2021}]{ref8}
Chengyu Zheng, Xiulian Peng, Yuan Zhang, Sriram Srinivasan, and Yan Lu.
\newblock Interactive speech and noise modeling for speech enhancement.
\newblock In {\em Proceedings of the AAAI Conference on Artificial
  Intelligence}, volume~35, pages 14549--14557, 2021.

\end{thebibliography}

\end{document}